\documentclass[11pt]{article}
\usepackage[margin=1in]{geometry}
\usepackage{amsmath,amssymb}
\usepackage{graphicx}
\usepackage[numbers,sort&compress]{natbib}
\usepackage[colorlinks=true,linkcolor=blue,citecolor=blue,urlcolor=blue]{hyperref}
\usepackage{url}
\DeclareMathOperator{\Tr}{Tr}

\title{Timeless Histories: Quantum Measurement and the Maximum Entropy Principle}
\author{Alexei V. Tkachenko\\[2pt]
\small Center for Functional Nanomaterials, Brookhaven National Laboratory, Upton, NY 11973, USA\\
\small \texttt{oleksiyt@bnl.gov}}
\date{\today}

\begin{document}
\maketitle

\begin{abstract}
The quantum measurement problem, i.e. the apparent conflict between unitary quantum evolution and non-unitary, stochastic wave-function collapse, remains unresolved a century after the formulation of quantum mechanics. We first review the standard picture, from the Copenhagen prescription with its Heisenberg cut, through von Neumann's movable cut, to environment-induced decoherence, which explains the emergence of stable classical records but still presupposes the Born rule and therefore cannot by itself replace the measurement postulates. We then connect this problem to the Maximum Entropy Principle in two complementary ways. First, decoherence drives the measurement apparatus toward the least biased state compatible with the dynamically protected pointer distribution, so that the emergent collapse may be viewed as a thermodynamic relaxation toward constrained maximum entropy. Second, we propose a Timeless Histories formulation of quantum mechanics in which the primitive objects are ordered sequences of events rather than evolving wave functions. Their conditional probabilities are assigned, relative to specified refinements, by a single Born-Boltzmann rule that combines the quadratic structure of Born's rule with Boltzmann's counting of equally probable microstates. The textbook form of Born's rule, the projection postulate, density operators, and wave functions are then recovered as derived informational constructs, while collapse becomes Bayesian conditioning on recorded events. Time and space enter only subsequently through unitary translation symmetry. Because the probability rule is postulated independently of any physical collapse or classical observer, decoherence is no longer asked to justify the rule used to interpret its own reduced states, and can be invoked to explain the approximate additivity and stability of macroscopic records, i.e. the emergence of classicality.
\end{abstract}

\section{Introduction}
\label{sec:intro}
The current formulation of Quantum Mechanics (QM) has just passed its centennial. Possibly the best way we physicists could celebrate this anniversary is by settling the conceptual controversies that still surround this remarkable theory. There is, of course, the standard formulation closely associated with the Copenhagen Interpretation. However, the very fact that the quantum theory has multiple {\it interpretations}, with no agreement in sight \cite{vonNeu, Bohm, Wigner, Everett, Wheeler, ManyWorlds, Deutsch1985, schlosshauer2007decoherence, Bell_1990, SCHLOSSHAUER2013surv}, stems from well-recognized issues that have remained unresolved since the early days of QM.

At the core of the disagreement is the quantum measurement process, which invokes the so-called Wave Function Collapse (WFC) to model an interaction between a quantum system and a classical apparatus. Since classical systems ultimately need to be described by QM as well, one arrives at a seemingly unsolvable problem: how can the non-unitary and non-deterministic measurement process be reduced to the unitary and deterministic quantum evolution? In this paper, we propose a way to remove the WFC from the set of QM postulates. The remaining measurement content is absorbed into a single probability axiom, which we call the Born--Boltzmann rule: it combines Born's rule with the Maximum Entropy Principle \cite{jaynes1957a,jaynes1957b}, and reduces to the Boltzmann-style counting of equally probable microstates in the classical limit. In this framework, the collapse becomes Bayesian updating of the available information rather than a physical process, which in turn allows one to employ the previously proposed mechanism of environment-induced decoherence to justify the quantum-to-classical transition \cite{zeh1970, Zurek_PT, Zurek_RMP, Hartle, schlosshauer2007decoherence} without logical circularity. Notably, the resulting formulation never refers to the wave function as a fundamental notion; as discussed below, this avoids the standard formulation of the notorious definite-outcome problem.

The Copenhagen formulation of QM, championed by N.~Bohr and W.~Heisenberg, implies that a quantum system has to coexist with a classical measurement apparatus and/or a classical observer. But even among the scientists mostly aligned with that school of thought, there was no unity about the nature of the boundary between the quantum and classical domains, also known as the Heisenberg cut. E.~Schr\"odinger introduced his famous Cat, in a letter to A.~Einstein, as part of a {\it gedankenexperiment} intended largely to ridicule the Copenhagen Interpretation and the role of the observer in it. Einstein, himself the most prominent opponent of the idea that the world is probabilistic at its most fundamental level, famously wrote in a letter to M.~Born: ``bin ich \"uberzeugt, da\ss{} der nicht w\"urfelt'' (``I am convinced that He does not roll dice''). He (Einstein, not God) hoped that the quantum uncertainty would one day emerge as a consequence of a theory involving yet unknown ``hidden variables''. A wide class of such {\it local} hidden-variable theories has by now been ruled out by the experimentally observed violation of Bell's inequality \cite{Bell, Bell1, Bell_exp72, Bell_exp82}; ironically, J.~Bell himself favored the explicitly nonlocal de Broglie--Bohm pilot-wave theory \cite{Bell_1990, Bohm}.

\begin{figure}[h!]
\centering
\includegraphics[width=0.64\linewidth]{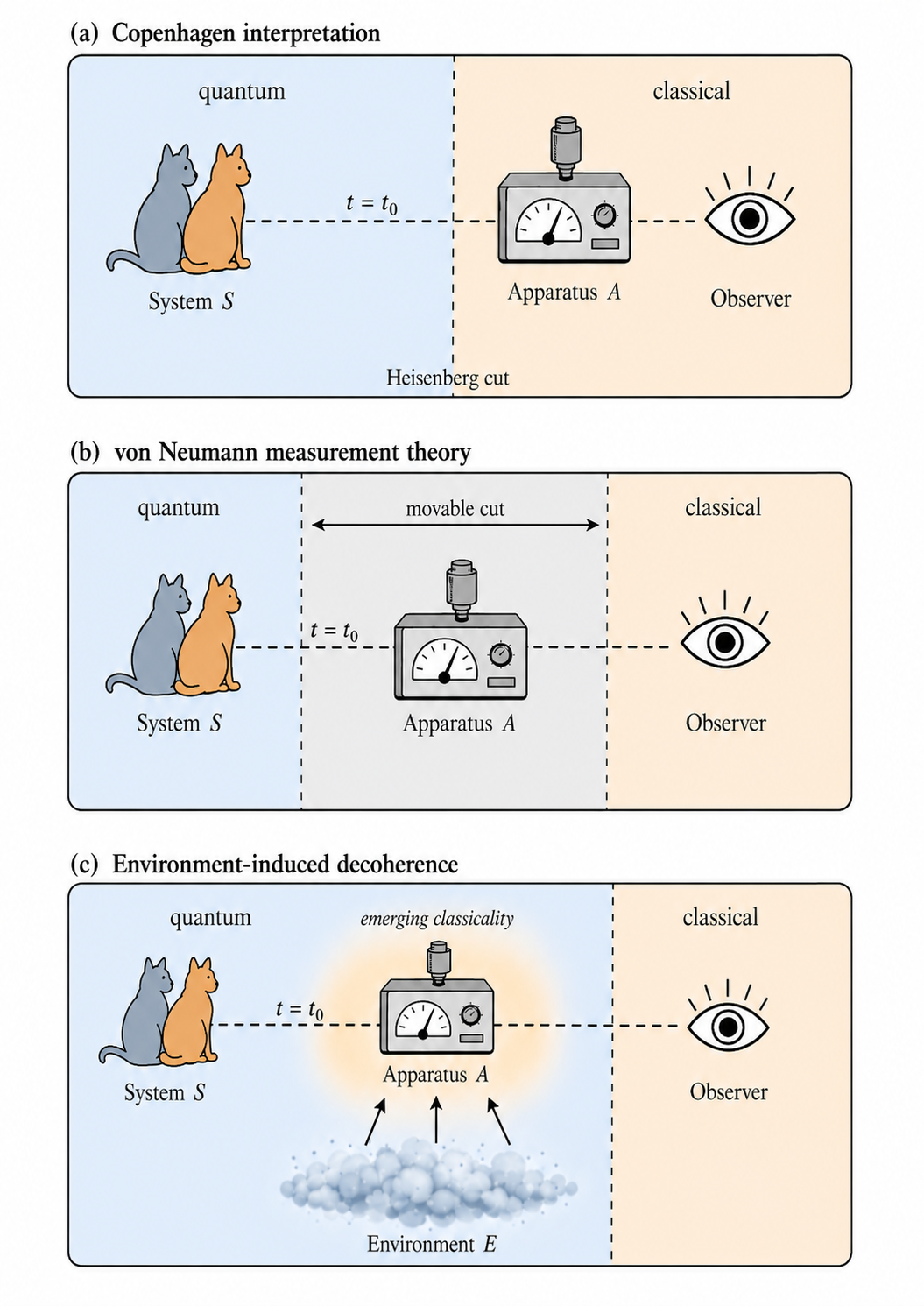}
\caption{The evolving picture of the quantum measurement. The System $\mathcal S$ is represented by Schr\"odinger's cat in a superposition of two states; the horizontal dashed connector marks the measurement interaction at $t=t_0$, while the vertical dashed lines mark the boundaries between the quantum and classical domains. (a) In the Copenhagen formulation, the quantum System coexists with a classical Apparatus and Observer; the Heisenberg cut must be drawn between the two domains, but its location is not specified by the theory. (b) In von Neumann's measurement theory, the Apparatus occupies a ``grey zone'': it can be equivalently treated as a quantum or a classical object, so the cut is movable, but not removable. (c) Within the environment-induced decoherence program, the Apparatus remains a part of the quantum domain, but the continuous monitoring by the Environment $\mathcal E$ einselects its pointer states and suppresses the coherence between them, endowing it with emerging classicality. Within the standard formulation, however, the boundary separating the classical Observer must still be drawn: the probabilistic interpretation of the decoherence program's tools rests on Born's rule, and hence on a classical domain (Sec.~\ref{sec:notenough}). In the formulation proposed in this paper, the cut is erased altogether (Secs.~\ref{sec:timeless}--\ref{sec:maxent_decoherence}).}
\label{fig:cut}
\end{figure}

The rest of the paper is organized as follows. Sec.~\ref{sec:statusquo} articulates the status quo, summarized pictorially in Fig.~\ref{fig:cut}: from the Copenhagen prescription and von Neumann's movable cut to environment-induced decoherence and its residual logical circularity. Sec.~\ref{sec:timeless} introduces the histories-based formulation: three postulates that refer neither to the collapse, nor to a classical observer, nor to the wave function itself, with Born's rule following as a special case of the probability postulate. Sec.~\ref{sec:recovering} recovers the standard formulation, from the measurement records to the density operators and wave functions, and introduces time and space through an additional translation-symmetry postulate. Sec.~\ref{sec:maxent_decoherence} revisits decoherence as a dynamical implementation of the same Maximum Entropy Principle, and Sec.~\ref{sec:discussion} concludes.

\section{Measurement postulates and the decoherence program}
\label{sec:statusquo}

\subsection{Copenhagen interpretation: quantum system and classical observer}
\label{sec:copenhagen}
A coherent mathematical formulation of QM is normally traced back to the books by P.~A.~M.~Dirac \cite{Dirac:1930:PQM} and J.~von Neumann \cite{vonNeu}. The state of a quantum system is described by a unit vector in a complex Hilbert space, known as the wave function $|\Psi\rangle$. Its time evolution is given by a deterministic unitary transformation, which can be expressed, e.g., in the form of the Schr\"odinger equation:
\begin{equation}
\label{Schro}
    i\hbar\frac{\partial\left| \Psi\right \rangle}{\partial t}= {\widehat H}\left| \Psi\right \rangle
\end{equation}
However, at the moment of measurement the system undergoes the WFC: a non-unitary jump onto one of several eigenstates $|\psi_j\rangle$ specific to that measurement,
\begin{equation}
\label{projection}
|\Psi\rangle \;\longrightarrow\; |\psi_j\rangle.
\end{equation}
This jump is non-deterministic, with the probabilities of different outcomes dictated by the projections of the wave function onto the corresponding eigenstate vectors:
\begin{equation}
\label{BornRule}
p_j=\left|\langle \Psi|\psi_j\rangle\right|^2,
\end{equation}
the property known as Born's rule. Together, Eqs.~(\ref{projection}) and (\ref{BornRule}) constitute the projection postulate, which supplements the unitary evolution, Eq.~(\ref{Schro}), as an independent axiom.

In the Copenhagen picture, the quantum system described by $|\Psi\rangle$ must coexist with a classical measurement apparatus and/or observer, which itself does not obey Eq.~(\ref{Schro}) and whose readings are treated as definite classical facts, as illustrated in Fig.~\ref{fig:cut}(a). The theory does not specify where the boundary between the two domains, the Heisenberg cut, should be placed; it only requires that it be placed somewhere. The internal consistency of this prescription therefore hinges on the statistics of measurement outcomes being insensitive to the location of the cut. It was von Neumann who turned this expectation into an explicit demonstration.

\subsection{Von Neumann's measurement theory}
\label{sec:vonNeumann}
Von Neumann \cite{vonNeu} distinguished two types of evolution of a quantum state: the unitary and deterministic dynamics generated by the Schr\"odinger equation (``Process 2''), and the non-unitary, stochastic transformation accompanying a measurement (``Process 1''). The latter formalizes the WFC: in the language of the density operator, Process 1 erases the off-diagonal elements in the basis specific to a particular measurement. Mathematically, this is equivalent to assigning probabilities to different outcomes according to Born's rule and discarding the quantum interference between them.

Von Neumann paid special attention to the quantum measurement process by explicitly introducing the measurement Apparatus into the quantum-mechanical description. In his scheme, the quantum Apparatus $\mathcal {A}$ gets entangled with the System of interest $\mathcal {S}$ through a unitary ``pre-measurement'' process:
\begin{equation}
\label{premeasure}
    \left(\sum_j c_j|\psi_j\rangle\right)|\alpha_0\rangle \rightarrow \sum_j c_j|\psi_j\rangle|\alpha_j\rangle
\end{equation}
Here $|\psi_j\rangle$ and $|\alpha_j\rangle$ represent the states of $\mathcal S$ and $\mathcal {A}$, respectively. After that, Process 1 applied to the Apparatus erases the off-diagonal terms in the density matrix of the combined $\mathcal {S+A}$ system:
\begin{equation}
\label{collapse}
\hat{\rho}^{SA}\rightarrow \sum_j \left|c_j\right|^2|\psi_j\rangle|\alpha_j\rangle\langle \alpha_j|\langle \psi_j|
\end{equation}
As noted above, such an irreversible and non-unitary transformation is not consistent with the Schr\"odinger equation. Crucially, however, the resulting statistics of outcomes is exactly the same as if Process 1 had been applied directly to $\mathcal S$, without ever including the Apparatus in the quantum description: in both cases, the outcome $j$ is observed with probability $|c_j|^2$. A quantum measuring apparatus is thus operationally equivalent to a classical one. By iterating this construction ($\mathcal S \rightarrow \mathcal A \rightarrow \mathcal A' \rightarrow \dots$, the von Neumann chain), one concludes that the interface between the classical and quantum domains can be shifted arbitrarily, without any effect on the observable predictions. As illustrated in Fig.~\ref{fig:cut}(b), the Apparatus occupies a ``grey zone'': it can be equivalently treated as a quantum or a classical object. In the spirit of the Copenhagen interpretation, von Neumann thus assumes that there is an interface between the two domains, but the Heisenberg cut is movable rather than fixed. Since nothing in the formalism localizes the collapse, von Neumann was ultimately led to place it at the level of the conscious observer, the last link of the chain \cite{vonNeu,wigner1995remarks}.

\subsection{Environment-induced decoherence and the origin of classicality}
\label{sec:EID}
An important development in understanding the nature of the quantum-to-classical transition was the decoherence paradigm, pioneered by H.~Zeh \cite{zeh1970} and brought to prominence in the 1990s thanks to the works by W.~Zurek and others \cite{Joos1985, Zurek_PT, Hartle, schlosshauer2007decoherence}. Within the decoherence program, classicality is no longer imposed through an external classical apparatus; instead, one considers a regular unitary evolution of an extended quantum system that contains, in addition to $\mathcal {S+A}$, another subsystem called the Bath or Environment ($\mathcal {E}$). As a starting point, consider the expanded $\mathcal {S+A+E}$ quantum system described by the following Hamiltonian:
\begin{equation}
\label{Hmodel}
\widehat{H} =\widehat{H}_S+\widehat V\delta (t-t_0) + \widehat{H}_{E}(\widehat A)
\end{equation}
Here $\widehat{H}_S$ is the Hamiltonian of $\mathcal S$, and $\widehat{V}$ is the coupling between $\mathcal S$ and $\mathcal {A}$ responsible for their entanglement, similar to Eq.~(\ref{premeasure}), at time $t=t_0$. The final term describes the Environment $\mathcal E$, which is coupled to the Apparatus only through a single operator $\widehat {A}$. That operator commutes with $\widehat{H}_S+\widehat{H}_E (\widehat A)$ and therefore represents a conserved quantity at any time except at $t=t_0$. The eigenstates $|k\rangle$ of $\widehat {A}$ are known as the pointer states of the Apparatus.

Interactions with the environment preserve the pointer states but destroy the coherence between them; Zurek termed this mechanism environment-induced superselection, or einselection \cite{Zurek_PT, Zurek_RMP}. This is a key observation within the decoherence paradigm, confirmed by a variety of model calculations and experiments \cite{caldeira1981, Joos1985, modelDecoherence, Zurek_RMP, OpenSystems, schlosshauer2007decoherence, SCHLOSSHAUER2019}. The standard argument \cite{Zurek_RMP, schlosshauer2007decoherence} proceeds as follows. Let the environment be initially in the state $|\epsilon_0\rangle$. Since the pointer projectors commute with the Hamiltonian (\ref{Hmodel}), the subsequent evolution is conditional on the pointer state: each sector $k$ drives the environment with its own Hamiltonian, $\widehat H_E(A_k)$, so that
\begin{equation}
\label{conditionalE}
|k\rangle|\epsilon_0\rangle \;\longrightarrow\; |k\rangle|\epsilon_k(t)\rangle,\qquad |\epsilon_k(t)\rangle=e^{-\frac{i}{\hbar}\widehat H_E(A_k)\,t}\,|\epsilon_0\rangle.
\end{equation}
In this way, the Environment continuously monitors the pointer observable and keeps a record of it. Taking the partial trace of the overall density operator over the environment then yields the reduced density operator of the Apparatus,
\begin{equation}
\label{dephasing}
\rho^A_{kk'}(t)=\rho^A_{kk'}(0)\,r_{kk'}(t),\qquad r_{kk'}(t)=\langle\epsilon_{k'}(t)|\epsilon_k(t)\rangle.
\end{equation}
The decoherence factor $r_{kk'}(t)$ is the overlap of the conditional environment states: the better the Environment distinguishes the pointer states $k$ and $k'$, the smaller the residual coherence between them. For a macroscopic environment with many independent degrees of freedom, this overlap is a product of many factors, each of magnitude below unity, so that $\left|r_{kk'}(t)\right|$ decays very fast, typically exponentially both in time and in the number of environmental degrees of freedom \cite{Joos1985, Zurek_RMP, schlosshauer2007decoherence}.

Thus, the decoherence process leads to the vanishing of the off-diagonal terms in the reduced density operator of the Apparatus, and therefore of the $\mathcal {S+A}$ subsystem as well. Importantly, this does not mean that the overall $\mathcal {S+A+E}$ system evolves towards one of the common eigenstates of the operator $\widehat A$, as that would be inconsistent with the unitary evolution. This evolution formally preserves the entropy associated with the overall density operator,
\begin{equation}
\label{entropy}
S=-\Tr\left(\hat\rho \ln \hat\rho\right).
\end{equation}

Due to the decoherence, the $\mathcal {S+A}$ subsystem becomes {\it asymptotically indistinguishable} from a classical mixture of eigenstates of the operator $\widehat {A}$, which justifies the application of von Neumann's Process 1, Eq.~(\ref{collapse}). In this sense, within the decoherence program the non-unitary Process 1 is no longer a fundamental postulate: it emerges as an effective, ``for all practical purposes'' description of the underlying unitary dynamics of the larger system. The extra entropy generated by Process 1 arises because we discard the information associated with the entanglement between the $\mathcal {S+A}$ subsystem and the Environment. Note that this argument is valid even if the $\mathcal E$ subsystem was originally in a pure rather than a mixed state.
\subsection{Quantum measurements and the Second Law}
The decohered state of the Apparatus admits a transparent thermodynamic characterization. In the model of Eq.~(\ref{Hmodel}), the Apparatus is coupled to the Environment only through the pointer observable $\widehat A=\sum_k A_k \widehat P_k$, which commutes with the Hamiltonian at all times except the moment of measurement. The distribution of the pointer variable is therefore a constant of motion: the probabilities
\begin{equation}
p_k=\Tr\left[\widehat P_k\,\widehat\rho^A\right]
\end{equation}
are protected by the conservation law, while all the remaining information carried by the state of the Apparatus, i.e. the coherences between different pointer sectors, is exposed to the Environment. The suppression of these coherences, Eq.~(\ref{dephasing}), drives the reduced density operator toward its block-diagonal (``pinched'') form,
\begin{equation}
\label{pinch}
\widehat\rho^A(t)\;\longrightarrow\;\widehat\rho^A_D=\sum_k \widehat P_k\, \widehat\rho^A\,\widehat P_k\,,
\end{equation}
which, for non-degenerate pointer states $|k\rangle$, reduces to $\widehat\rho^A_D=\sum_k p_k|k\rangle\langle k|$. The decohered state (\ref{pinch}) admits a simple variational characterization: it is the unique maximizer of the von Neumann entropy, Eq.~(\ref{entropy}), among all the states consistent with the conserved pointer distribution:
\begin{equation}
\label{maxent_var}
\widehat\rho^A_D=\underset{\widehat\rho}{\arg\max}\left\{\,S(\widehat\rho)\ \Big|\ \Tr[\widehat P_k\widehat\rho\,]=p_k\ \ \forall k\right\}
\end{equation}
Indeed, maximizing $S(\widehat\rho)=-\Tr[\widehat\rho\ln\widehat\rho\,]$ with Lagrange multipliers enforcing the constraints yields $\widehat\rho\propto\exp\big(\sum_k\lambda_k\widehat P_k\big)$, which is diagonal in the pointer basis; matching the constraints then gives $\sum_k p_k|k\rangle\langle k|$. This establishes Eq.~(\ref{maxent_var}) for non-degenerate pointer projectors. If the pointer sectors are degenerate, the pinched state (\ref{pinch}) preserves the state within each sector, and it maximizes the entropy only when the dynamically protected intra-sector information is included among the constraints; if the sector probabilities $p_k$ alone are fixed, the maximizer is instead $\sum_k (p_k/d_k)\widehat P_k$, with $d_k$ the rank of $\widehat P_k$. In either case, the decohered state is the least informative one compatible with what the dynamics protects. Equivalently, for an arbitrary state one has the identity
\begin{equation}
\label{coherence}
S\!\left(\widehat\rho^A_D\right)=S\!\left(\widehat\rho^A\right)+S\!\left(\widehat\rho^A\,\big\|\,\widehat\rho^A_D\right),
\end{equation}
where $S(\widehat\rho\|\widehat\sigma)=\Tr[\widehat\rho\,(\ln\widehat\rho-\ln\widehat\sigma)]\ge 0$ is the relative entropy, known in this context as the relative entropy of coherence \cite{baumgratz2014}. Thus the decoherence can only increase the entropy of the Apparatus, and the entropy gain is exactly the information content of the discarded coherences. The Environment thus acts as a dynamical enforcer of Jaynes' Maximum Entropy Principle \cite{jaynes1957a,jaynes1957b}: it suppresses the coherence information that is not protected by the pointer-sector structure, driving the state of the Apparatus toward the least biased form compatible with the conserved pointer distribution. This gives the einselection of the pointer basis a transparent thermodynamic meaning: the preferred basis is singled out by the conservation law, and everything else is randomized; the pinched state is the least informative state compatible with the dynamically protected pointer structure, loosely analogous to the microcanonical equipartition.

From this perspective, the emergent Process 1 can be viewed as a special case of the second law of thermodynamics: in the decoherence limit of a large, effectively Markovian environment, the entropy of the open Apparatus increases toward the constrained maximum, Eq.~(\ref{maxent_var}). (For a finite or strongly non-Markovian environment, transient recoherence and entropy decrease are possible; Eq.~(\ref{pinch}) then describes the asymptotic, coarse-grained state rather than a monotonic trajectory.) No entropy is created out of nothing: the overall $\mathcal{S+A+E}$ evolution remains unitary, and the growth of the reduced entropy $S(\widehat\rho^A)$ reflects the buildup of entanglement between the Apparatus and the Environment, rather than any fundamental irreversibility. The second law itself is, of course, an emergent phenomenon. Its modern derivations from the underlying unitary dynamics rely on the typicality of maximum-entropy states in large Hilbert spaces, established in the quantum typicality program of Goldstein, Lebowitz and collaborators \cite{thermal2, goldstein2010} and, independently, by Popescu, Short and Winter \cite{popescu2006}, as well as on the eigenstate thermalization hypothesis \cite{quantum_statmech, srednicki1994, rigol2008, thermal1}. The environment-induced erasure of the pointer coherences is one more instance of the same relaxation toward the maximum-entropy state compatible with the conservation laws.

Pictorially, the continuous monitoring by the Environment endows the Apparatus with emerging classicality, as shown in Fig.~\ref{fig:cut}(c): the Apparatus remains a quantum system, yet becomes indistinguishable from a classical one.

\subsection{Why decoherence is not enough}
\label{sec:notenough}
In light of the decoherence paradigm, it would be tempting to simply drop the WFC out of the formulation of QM. However, that would not resolve the original issue of inconsistency, because of the concern of logical circularity, articulated by Schlosshauer and others \cite{schlosshauer2007decoherence, Schloss_decoh, ADLER2003135} and widely acknowledged within the decoherence program itself. Indeed, while the partial trace is definable as a purely mathematical operation, its physical reading within the decoherence program relies on Born's rule: the reduced density operator is interpreted precisely as the object that reproduces, via Born's rule, the correct probabilities for all measurements performed on the subsystem. Within the textbook formulation, those measurements are in turn described by the projection postulate. In other words, decoherence explains the dynamical suppression of interference and the stability of pointer records, but does not independently supply the probability rule, or the conditional state-update rule, needed to interpret the measurement outcomes: it presupposes the very measurement postulates it is meant to explain. It therefore cannot, on its own, replace those postulates; within the standard, collapse-based formulation, the boundary between the quantum and the classical domains must still be drawn somewhere, as emphasized in Fig.~\ref{fig:cut}(c). For this reason, the decoherence program is currently viewed as an important consistency check for QM, rather than a major revision of its logical structure \cite{SCHLOSSHAUER2019}.

The circularity concern has been recognized within the decoherence program itself, and Zurek's program of deriving Born's rule from entanglement-assisted invariance, or ``envariance'' \cite{Zurek_PRA2005, Zurek2018, SebensCarroll}, was developed precisely in order to address it, by seeking a route to the probabilities that does not presuppose them. Other proposals for deriving the measurement rules rather than postulating them \cite{vaidman2020derivations} include decision-theoretic arguments within the Many-Worlds Interpretation (MWI) \cite{Everett, Wheeler, deutsch1999quantum, ManyWorlds} and, on the mathematical side, Gleason-type uniqueness theorems \cite{Gleason1957, galley2017classification, Masanes2019}. The assumptions behind the former two have been extensively debated \cite{Barnum2000, envience2023noBorne}, while the latter establish the uniqueness of the Born measure but tacitly rely on the measurement postulates for its operational meaning. We return to these approaches in Sec.~\ref{sec:discussion}.

One of the directions that has the potential of avoiding the measurement problem altogether is the Consistent Histories (CH) formulation introduced in the works of R.~B.~Griffiths \cite{hist_grif}, further developed by R.~Omn\`es \cite{hist_omnes}, and by M.~Gell-Mann and J.~B.~Hartle as the Decoherent Histories QM \cite{Hartle, decoh_hist}. By assigning weights directly to coarse-grained quantum trajectories (histories), these approaches remove the WFC from the list of the fundamental notions, and in our view they contain the essential ingredients of the solution. Gell-Mann and Hartle, in particular, explicitly sought to have the mutual exclusivity of the macroscopic alternatives emerge from the system dynamics, i.e. from decoherence. In the end, however, both versions retain the consistency (decoherence) condition as a separate requirement imposed on the admissible families of histories, and both take the weight assignment, itself a generalization of Born's rule, as primitive. In that sense, the programs reformulate the measurement problem in a very promising language, but stop short of closing the loop. A complementary route was explored in Ref.~\cite{Tkachenko2020}, where the branching of mutually exclusive classical histories, and with it the consistency property, was shown to emerge within conventional QM, conditional on the informational isolation of the measurement records. The formulation developed below can be viewed as an attempt to complete the program initiated by Griffiths: to keep the histories and their weights, to dispense with the consistency condition as an axiom, and to let einselection supply what it was meant to supply all along.

\section{Timeless Histories formulation of QM}
\label{sec:timeless}
Below, we present a set of QM postulates that is a streamlined variation of the histories-based approaches \cite{hist_grif, hist_omnes, Hartle, decoh_hist, hist_Isham}. Importantly, and in contrast to the original Griffiths formulation, the consistency condition is not among them: as discussed in Sec.~\ref{sec:maxent_decoherence}, its role is taken over by decoherence. In addition, perhaps paradoxically, in our construction the quantum histories are introduced without any explicit mention of time or temporal evolution: a history is simply an ordered sequence of events, and it is the ordering, rather than a time parameter, that plays the fundamental role. Time, along with space, is introduced afterwards (Sec.~\ref{sec:spacetime}), through an additional symmetry postulate that associates unitary transformations with translations; the probabilistic core of the theory, Postulates 1--3, does not depend on it. We stress that physical time is not being derived from the ordering: once time labels are available, the ordinal order of a temporal history is simply set consistently with the temporal one. The claim is only that the probability rule itself requires nothing beyond the ordering. Equally importantly, the postulates below never refer to the wave function, or to any ``state of the system'' as a fundamental notion: the primitive elements of the description are the events and their probabilities, while the quantum states emerge as derived, informational constructs (Sec.~\ref{sec:density}).

\begin{enumerate}
    \item {\bf Hilbert Space:} {\it An isolated physical system is associated with a complex Hilbert space ${\mathcal H}$, conventionally referred to as its state space. The state space of a composite system is the tensor product of the Hilbert spaces associated with the component systems: ${\mathcal H}={\mathcal H_1}\otimes{\mathcal H_2}$.}
    \item {\bf Physical Events:} {\it A physical observation is characterized by a sample space $\Omega$ of mutually exclusive elementary outcomes. The events, i.e. the subsets $\omega \subseteq \Omega$, are represented by mutually commuting projection operators $\widehat P(\omega)$ acting in ${\mathcal H}$, in such a way that the logical operations are preserved:
\begin{align}
\widehat P(\omega \wedge \omega')&=\widehat P(\omega) \widehat P( \omega')\\
\widehat P(\omega \vee \omega')&=\widehat P(\omega) +\widehat P( \omega')-\widehat P(\omega) \widehat P( \omega')\\
\widehat  P(\varnothing)=0,&\qquad \widehat  P(\Omega)=\widehat I
\end{align}
}
\end{enumerate}
In other words, the events of a given sample space form a Boolean algebra, which Postulate 2 maps isomorphically onto an algebra of mutually commuting projectors. In particular, the negation of an event corresponds to the complementary projector, $\widehat P(\Omega \backslash \omega)=\widehat I -\widehat P(\omega)$.

Postulate 2 is conventionally formulated by associating a physical observable $A$ with a Hermitian operator $\widehat {A}$ in space ${\mathcal H}$. To recover that formulation, consider a sample space $\Omega$ made of all possible values $A_k$ of that observable. There is a one-to-one correspondence between each $A_k$ and the projection operator onto the respective eigenspace, $\widehat P^A_k$. The operator $\widehat {A}$ can be represented as a linear combination of these mutually orthogonal projections:
 \begin{equation}
    \widehat {A}= \sum_k A_k\widehat P^A_k
\end{equation}
Note that the set of projection operators $\widehat P^A_k$ represents a Projective Decomposition of the Identity (PDI):
\begin{align}
\widehat P^A_k\widehat P^A_j&=\widehat P^A_k\delta_{kj}\\
\sum_k \widehat P_k^A&= \widehat I
\end{align}
Crucially, while the projectors representing events from the same sample space commute with each other by construction, the projectors associated with two different sample spaces, $\Omega_1$ and $\Omega_2$, need not. This is the only, yet essential, departure from classical probability theory, and the ultimate source of all quantum phenomena.

The sample space $\Omega$ introduced in Postulate 2, and the collection of its subsets called events, are fundamental objects of probability theory. The last missing element is the probability itself. As we will see, rather than a single additive measure on the event space, it is introduced as a rule that assigns normalized conditional probabilities to the alternatives within a history. Before formulating this rule, we define a history as an ordered sequence of events, $\vec \omega^{(n)}=(\omega_1,...,\omega_{n})$, drawn from the respective sample spaces $\Omega_1,\dots,\Omega_n$; equivalently, it is a subset of the composite sample space, $\vec \omega^{(n)} \subseteq \Omega_1 \times ... \times \Omega_n$. We emphasize that the sample spaces $\Omega_i$ are in general mutually incompatible, i.e. represented by non-commuting sets of projectors. We also stress that the index $i$ ordering the events is, at this stage, purely ordinal: it does not refer to any time parameter.

\begin{enumerate}
\setcounter{enumi}{2}
\item {\bf Probability Rule (Born--Boltzmann):}
{\it Each history is assigned a non-negative weight
\begin{equation}
\label{weight}
W(\vec\omega^{(n)})=\Tr \left[\widehat C^\dagger(\vec \omega^{(n)}) \widehat C(\vec \omega^{(n)}) \right],
\end{equation}
where $\widehat C(\vec \omega^{(n)})$ is the Chain Operator of the history:
\begin{equation}
\label{chain}
\widehat C(\vec \omega^{(n)})=\widehat P(\omega_n)\widehat P(\omega_{n-1})\cdots\widehat P(\omega_1).
\end{equation}
Let the history be split as $\vec\omega^{(n)}=(L,\omega_k,R)$, where the sub-histories $L$ and $R$ consist of the events preceding and following $\omega_k\in\Omega_k$, respectively. The conditional probability of the event $\omega_k$, given the rest of the history and relative to the binary refinement $\{\omega_k,\bar\omega_k\}$, is
\begin{equation}
\label{probrule}
  \mathbb P(\omega_k| L,R;\mathcal M_k) =\frac{W(L,\omega_k,R)}{W(L,\omega_k,R)+W(L,\bar\omega_k,R)},
\end{equation}
where $\bar\omega_k=\Omega_k\backslash\omega_k$ is the complementary event, and $\mathcal M_k=\{\omega_k,\bar\omega_k\}$ indicates the binary alternative with respect to which the probability is defined.}
\end{enumerate}
The two alternatives $h_\omega=(L,\omega_k,R)$ and $h_{\bar\omega}=(L,\bar\omega_k,R)$ constitute a {\it binary refinement} of the coarser history. Since they exhaust the possibilities within the specified binary context $\mathcal M_k$, the rule is normalized by construction, $\mathbb P(\omega_k|L,R;\mathcal M_k)+\mathbb P(\bar\omega_k|L,R;\mathcal M_k)=1$, for an arbitrary event of arbitrary rank, at an arbitrary position within the history (the rule applies whenever the denominator is nonzero; its extension to a refinement with several mutually exclusive alternatives is immediate, with the denominator replaced by the sum of the corresponding weights). Below we suppress the label $\mathcal M_k$ whenever the refinement is clear from the context. We stress that the probability refers to the specified refinement: resolving the same coarse alternative differently, e.g. by a finer-grained interior measurement followed by discarding a part of the outcome, corresponds to a different set of histories and, in general, to different probabilities. As discussed below, this is the expected quantum contextuality rather than a defect of the rule. Note that the certain event $\Omega_k$ is represented by the identity operator, so inserting it into a history changes nothing: $(L,\Omega_k,R)=(L,R)$, and hence $W(L,\Omega_k,R)=W(L,R)$. Importantly, while the chain operators of the refinement add up to that of the coarse history, $\widehat C(L,R)=\widehat C(h_\omega)+\widehat C(h_{\bar\omega})$, their quadratic weights do not:
\begin{align}
\label{interference}
W(L,R)&=W(h_\omega)+W(h_{\bar\omega})+2\,{\rm Re}\, D(h_\omega,h_{\bar\omega}),\nonumber\\
D(h,h')&\equiv\Tr\left[\widehat C^\dagger_{h}\,\widehat C_{h'}\right].
\end{align}
The cross term, expressed through the interference functional $D$, represents the quantum interference between the two alternatives. As shown below, it vanishes in the classical (compatible) regime, as well as for the events located at the edges of a history; in general, however, the weight of a coherent coarse history differs from the sum of the weights of its refinements. This is a feature, not a bug: the chain operators add under logical coarse-graining, while the probabilities, being quadratic in them, capture the interference.

\subsection{The Born--Boltzmann rule and its implications}
\label{sec:BB}
The postulate introduced above has a number of noteworthy implications, which we now spell out.

\begin{figure}[t!]
\centering
\includegraphics[width=0.72\linewidth]{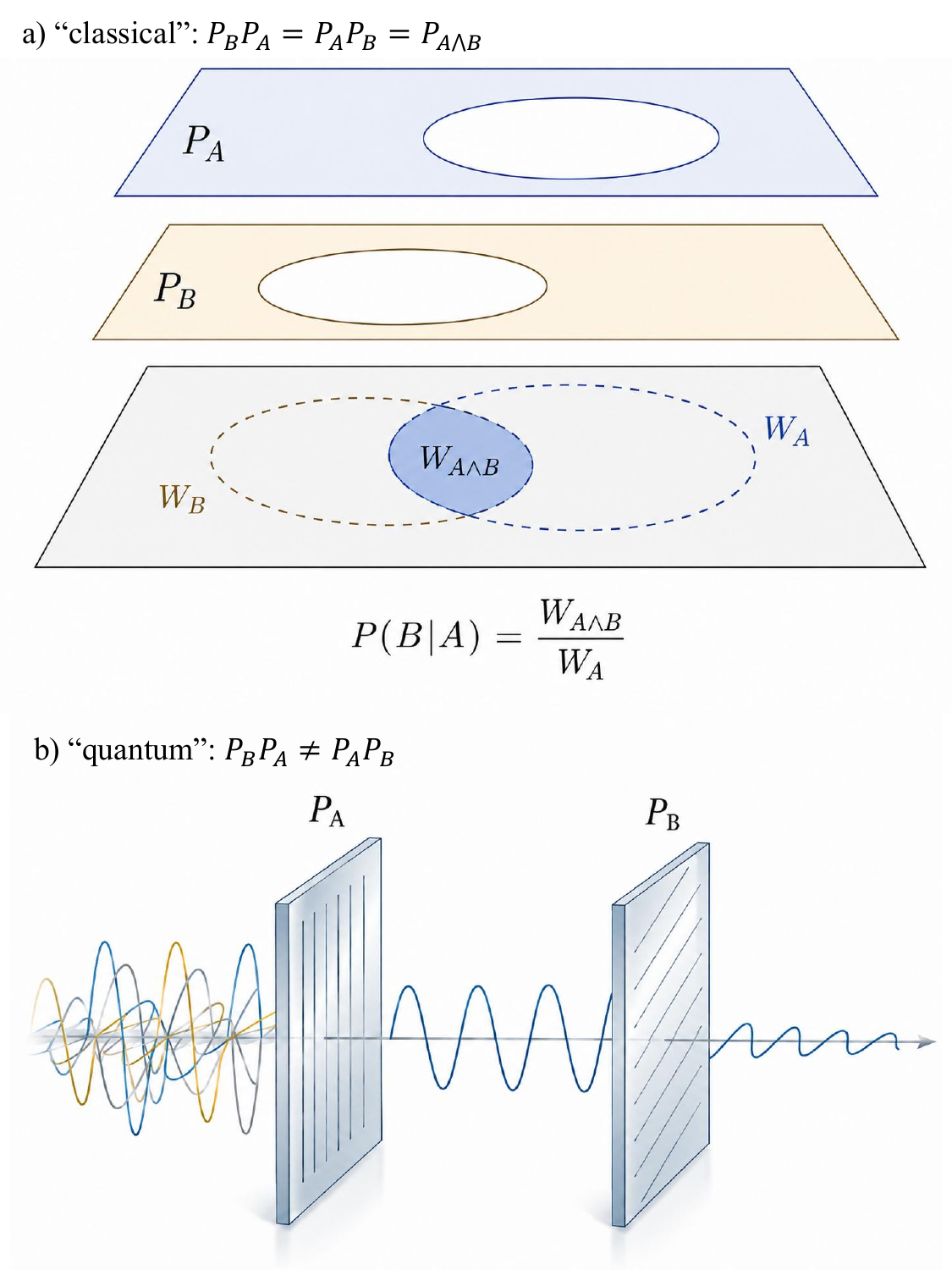}
\caption{The Born--Boltzmann rule at work. (a) The ``classical'' regime: two masks with aligned openings represent commuting projectors, whose product is itself a projector, $\widehat P_B\widehat P_A=\widehat P_A\widehat P_B=\widehat P_{A\wedge B}$. The weights count the microstates compatible with the respective events, and the conditional probability, $\mathbb P(B|A)=W_{A\wedge B}/W_A$, is the fraction of the state space compatible with both. (b) The ``quantum'' regime: unpolarized light, i.e. an ensemble of photons with maximally mixed polarization, described by the uniform maximum-entropy prior, passes through two non-aligned polarizers, which act as non-commuting projection operators; this is arguably the simplest quantum history. The transmitted intensity is proportional to the weight $W$, and the intensity transmitted by the second polarizer, relative to that emerging from the first, gives the conditional probability $W(\psi,\phi)/W(\psi)=\cos^2\theta$: Malus' law is Born's rule, Eq.~(\ref{BornBB}). Note that if the direction of the light is reversed, from right to left, the transmitted intensity, and hence the weight, stays the same, illustrating the order-reversal symmetry of the weight, Eq.~(\ref{weight}).}
\label{fig:BB}
\end{figure}

{\it (i) Boltzmann counting in the classical limit.} The physical content of the weight, Eq.~(\ref{weight}), is most transparent in the classical limit, when all the events of the history are mutually compatible, i.e. represented by commuting projectors. In that case, the product of the projectors is itself a projector, onto the intersection of the corresponding subspaces, $\widehat C^\dagger\widehat C=\widehat C=\widehat P(\omega_1\wedge\dots\wedge\omega_n)$, and the weight
\begin{equation}
\label{Wclassical}
W(\vec\omega^{(n)})=\Tr\, \widehat P(\omega_1\wedge\dots\wedge\omega_n)
\end{equation}
is an integer: it literally counts the number of mutually orthogonal ``microstates'' compatible with all the events of the history.\footnote{The present formulation is stated for finite-dimensional state spaces. Its extension to the infinite-dimensional case requires a trace-class reference state or a controlled limiting procedure, and is left for future work.} In the compatible case the interference functional vanishes, and the sum in the denominator of Eq.~(\ref{probrule}) coincides with the full weight, $W(L,\omega_k,R)+W(L,\bar\omega_k,R)=W(L,R)$. The probability rule (\ref{probrule}) then reduces to a ratio of two microstate counts. For instance, for a two-event history,
\begin{equation}
\mathbb P(\omega_B|\omega_A)=\frac{W(\omega_A\wedge \omega_B)}{W(\omega_A)}
\end{equation}
is literally the fraction of the state space compatible with the observation $\omega_B$, within the subspace singled out by the a priori information $\omega_A$, as illustrated in Fig.~\ref{fig:BB}(a). In other words, in the compatible case Postulate 3 reduces to classical probability theory supplemented with the Boltzmann-style postulate of equal a priori probabilities \cite{jaynes1957a,jaynes1957b}: every microstate consistent with the available information carries the same statistical weight. In particular, conditioning on a single event $X$ prepares the equal-weight mixture $\widehat P_X/\Tr\widehat P_X$ on the compatible subspace; if $X$ specifies the value of a macroscopic conserved quantity, e.g. the energy, to within a narrow window, this is precisely the microcanonical ensemble. The equal-a-priori-probability postulate of statistical mechanics is therefore not an independent assumption of our framework, but a corollary of Postulate 3. It is amusing to note that the weight $W$ of a compatible history coincides with Boltzmann's number of complexions, so that its logarithm plays the role of the Boltzmann entropy, $S=k_B\ln W$.

{\it (ii) Consecutive histories.} Consider an event $\omega$ located at the very edge of a history, say immediately preceding a sub-history $R$. Using the idempotency of the projectors and the cyclic property of the trace, $W(\omega,R)=\Tr[\widehat P(\omega)\widehat C^\dagger(R)\widehat C(R)\widehat P(\omega)]=\Tr[\widehat C^\dagger(R)\widehat C(R)\widehat P(\omega)]$; equivalently, the interference functional of an edge refinement vanishes identically, $D\left((\omega,R),(\bar\omega,R)\right)=\Tr[\widehat P(\omega)\widehat C^\dagger(R)\widehat C(R)\widehat P(\bar\omega)]=0$, since $\widehat P(\bar\omega)\widehat P(\omega)=0$. The weights of the edge events are therefore exactly additive, $W(\omega,R)+W(\bar\omega,R)=W(R)$, and the postulate (\ref{probrule}) yields $\mathbb P(\omega|R)=W(\omega,R)/W(R)$. Applying this result repeatedly, in accordance with the chain rule of conditional probabilities, one obtains the conditional probabilities relating two consecutive sub-histories:
\begin{equation}
\label{blockcond}
\mathbb P(L|R)=\frac{W(L,R)}{W(R)},\qquad \mathbb P(R|L)=\frac{W(L,R)}{W(L)}.
\end{equation}
For instance, for $L=(\omega_1,\omega_2)$: $\mathbb P(L|R)=\mathbb P(\omega_1|\omega_2\wedge R)\,\mathbb P(\omega_2|R)=W(L,R)/W(R)$.

{\it (iii) Prediction and retrodiction.} As a special case of Eq.~(\ref{blockcond}), the probability of a single event $\omega$ following a given history $\vec\omega$, and that of an event preceding it, are
\begin{align}
\mathbb P(\omega|\vec\omega)&=\frac{W(\vec\omega,\omega)}{W(\vec\omega)}\quad\text{(prediction)},\nonumber\\
\mathbb P(\omega|\vec\omega)&=\frac{W(\omega,\vec\omega)}{W(\vec\omega)}\quad\text{(retrodiction)}.
\label{predret}
\end{align}
For an interior event, the rule (\ref{probrule}) likewise reproduces the structure of the time-symmetric pre- and post-selection probabilities of Aharonov, Bergmann, and Lebowitz \cite{ABL1964}. Note that reversing the order of the entire sequence replaces $\widehat C$ with $\widehat C^\dagger$ and leaves the weight, Eq.~(\ref{weight}), unchanged; the predictive and retrodictive inferences are therefore governed by identical rules, in line with the timeless spirit of the construction. We stress that this is an ordinal reversal symmetry of the weight, not yet the dynamical time-reversal invariance, which involves an antiunitary transformation and is a property of specific Hamiltonians.

{\it (iv) Born's rule.} For rank-one events, Eq.~(\ref{predret}) takes the familiar form of Born's rule. Let the events $\psi$ and $\phi$ be represented by the projectors $\widehat P_\psi=|\psi\rangle\langle\psi|$ and $\widehat P_\phi=|\phi\rangle\langle\phi|$; then
\begin{equation}
\label{BornBB}
\mathbb P(\phi|\psi)=\frac{W(\psi,\phi)}{W(\psi)}=\frac{\Tr\left[\widehat P_\psi\widehat P_\phi\widehat P_\psi\right]}{\Tr \widehat P_\psi}=\left|\langle\phi|\psi\rangle\right|^2,
\end{equation}
and, by the time symmetry noted above, the corresponding retrodictive probability is the same. The simplest physical realization is shown in Fig.~\ref{fig:BB}(b): unpolarized light (equivalently, an ensemble of identically prepared photons), described by the uniform prior, passes through two non-aligned polarizers, with the transmitted intensity proportional to the weight. The intensity transmitted by the second polarizer, relative to that emerging from the first, gives the conditional probability $W(\psi,\phi)/W(\psi)=\cos^2\theta$: Malus' law is precisely Eq.~(\ref{BornBB}). The reversal of the light direction leaves the transmitted intensity unchanged. The operational content of this formula, i.e. its relation to the actual measurements performed by macroscopic devices, is established in Sec.~\ref{sec:born}.

In the general case, the projectors representing different events do not commute; their ordering then matters, and it is fixed by the ordering of the events within the history. The operator $\widehat C^\dagger\widehat C$ is no longer a projector, and its trace is no longer an integer: the counting of microstates is replaced by the weight, Eq.~(\ref{weight}), allowing for quantum interference between incompatible events. This is why we refer to Eq.~(\ref{probrule}) as the Born--Boltzmann rule: it combines, in a single axiom, the equal-a-priori-probability postulate of statistical mechanics (a uniform, maximum-entropy prior over the state space, together with Bayesian conditioning on the recorded events) with the quadratic amplitude structure of Born's rule, which is built into the noncommutative weight, Eq.~(\ref{BornBB}). We stress that the latter is genuinely postulated rather than derived: Eq.~(\ref{weight}) is not the unique noncommutative extension of the microstate counting (for instance, $W_\alpha=\Tr[(\widehat C^\dagger\widehat C)^\alpha]$ reduces to the same counting in the compatible limit for any $\alpha$), and we do not claim a uniqueness theorem here. It is worth noting, however, that the exact endpoint additivity derived in (ii) above is not generally satisfied for $\alpha\neq 1$, which singles out the trace weight within this family. The particular choice (\ref{weight}) is singled out by its consistency properties, by Gleason-type arguments \cite{Gleason1957, galley2017classification, Masanes2019}, by the formal equivalence with the standard formulation established in Sec.~\ref{sec:equivalence}, and, ultimately, by its agreement with experiment. An essentially identical forward-conditioning formula appears in the history-operator formulation of Castellani \cite{castellani2019}; the present construction differs in the Maximum-Entropy reading of the trace, in the treatment of measurements via record projectors (Sec.~\ref{sec:born}), and in the treatment of preparation: no density operator is supplied as an independent primitive; in the absence of any information, the a priori density operator is proportional to unity, and all preparation information is specified through conditioning events.

What the postulates do {\it not} contain is the WFC: no non-unitary, stochastic transformation of a state is ever invoked and, as shown in Sec.~\ref{sec:bayesian}, the projection postulate reappears as a theorem about conditional probabilities. Moreover, the postulates never refer to the wave function: the primitive elements of the theory are the events and the probabilistic relations among them. This avoids the standard formulation of the definite-outcome problem, namely the selection of a single actual branch out of a universally evolving quantum state: in the present framework, there is no such state to begin with. To be sure, the occurrence of events remains a primitive notion, just as it is in classical probability theory; what is claimed is not that stochasticity has been explained away, but that no additional selection mechanism, acting on top of a deterministic state evolution, needs to be invoked. That a particular event occurs, with the probability prescribed by Postulate 3, is then no more (and no less) mysterious than a die showing a particular face.

\section{Recovering standard QM}
\label{sec:recovering}

\subsection{Density operator and wave function}
\label{sec:density}
We start by showing how the density operator, and with it the notion of a quantum state, arises as a derived object. Suppose the events of a history $h$ are known to have occurred; we will refer to $h$ as the {\it prehistory}. According to Eq.~(\ref{blockcond}), the probability of a subsequent event $\omega$ is $\mathbb P(\omega|h)=W(h,\omega)/W(h)$; using the idempotency of the projectors and the cyclic property of the trace, it can be brought to the familiar form
\begin{equation}
\label{rhoh}
\mathbb P(\omega|h)=\Tr\left[\widehat P(\omega)\,\widehat\rho_h\right],
\qquad
\widehat\rho_h\equiv\frac{\widehat C(h)\,\widehat C^\dagger(h)}{W(h)}.
\end{equation}
The operator $\widehat\rho_h$ is Hermitian, positive semi-definite, and has a unit trace: it is the density operator induced by the prehistory, encoding everything the recorded past implies about the subsequent events. In the limiting case of an empty prehistory, it reduces to the uniform prior, $\widehat\rho\propto\widehat I$; for a single preparation event $X$, it is the maximally mixed state on the corresponding subspace, $\widehat P_X/\Tr\widehat P_X$ -- the least biased assignment consistent with the available information, in accordance with the Maximum Entropy Principle. One may say that conditioning on the events prepares the state ``out of unity''. Note also that appending one more event to the prehistory transforms the chain operator as $\widehat C\rightarrow\widehat P(\omega)\widehat C$ and, consequently, updates the state as $\widehat\rho_h\rightarrow\widehat P(\omega)\widehat\rho_h\widehat P(\omega)/\Tr[\widehat P(\omega)\widehat\rho_h\widehat P(\omega)]$; as discussed in Sec.~\ref{sec:bayesian}, this is nothing but L\"uders' rule. Finally, by the order-reversal symmetry of the weight, the conjugate combination $\widehat C^\dagger(h)\widehat C(h)/W(h)$ plays the very same role for the events {\it preceding} the prehistory: the two operators are the predictive and the retrodictive states induced by the same set of events.

More generally, the available information may involve classical uncertainty about the prehistory itself: let probabilities $f(h)$ be assigned to a set of mutually exclusive alternative prehistories $h$ (e.g., differing in the outcome of at least one recorded event), with $\sum_h f(h)=1$. The corresponding density operator is the classical mixture of the respective prehistory states,
\begin{equation}
\label{rhodef}
\widehat \rho \equiv \sum_h f(h)\,\widehat\rho_h=\sum_h f(h)\,\frac{\widehat C(h)\,\widehat C^\dagger(h)}{W(h)},
\end{equation}
and averaging Eq.~(\ref{rhoh}) over the prehistories yields, for an arbitrary subsequent event $\omega$, the standard expression
\begin{equation}
\label{probrho}
\mathbb P(\omega)=\Tr\left[\widehat P(\omega)\widehat\rho\right]
\end{equation}
Combining this with the spectral representation of a Hermitian operator, we recover the usual expression for its expected value:
\begin{equation}
   \langle  \widehat A \rangle=\sum_k A_k\, \mathbb P(a_k)= \Tr \left(\widehat A \widehat \rho \right)
\end{equation}

It is at this point that the notion of a quantum state, absent from the postulates, can be introduced. Let $|\Psi_i\rangle$ be the unit eigenvectors of $\widehat \rho$, and $p_i$ the corresponding eigenvalues; the spectral decomposition of the density operator then reads
\begin{equation}
    \widehat\rho=\sum_i p_i |\Psi_i\rangle\langle\Psi_i|
\end{equation}
The rank-one operators $|\Psi_i\rangle\langle\Psi_i|$ are the extremal points of the convex set of density operators; they define the pure states of the system, each represented by a unit vector, the wave function $|\Psi_i\rangle$. The wave function {\it as a state of preparation and evolution} is thus a derived, informational representation rather than a primitive ontic object (the rays themselves are, of course, already present in the projector geometry of Postulates 1 and 2). The non-negative factors $p_i$ represent classical probabilities, and any expected value is obtained by the ensemble averaging over the pure components with the weights $p_i$. The decomposition is not unique (e.g. for degenerate $p_i$), but all convex decompositions of a given $\widehat\rho$ lead to identical predictions.

\subsection{Measurement records and Bayesian updating}
\label{sec:born}
\label{sec:bayesian}
Born's rule has already been obtained in Sec.~\ref{sec:BB}, as a direct implication of the probability postulate. In the laboratory, however, the conditioning events are not abstract rank-one projectors, but readings of macroscopic devices. The statement ``the Apparatus $\mathcal A$ has measured $\mathcal S$ in the basis $|\psi_j\rangle$ and obtained the outcome $j$'' corresponds to the event $a_j$, represented by the {\it joint record--correlation} projector
\begin{equation}
\label{recordA}
\widehat E^A_j=|\psi_j\rangle\langle\psi_j|\otimes|\alpha_j\rangle\langle\alpha_j|,
\end{equation}
where $|\alpha_j\rangle$ are the pointer states of $\mathcal A$, and an identity factor over all remaining degrees of freedom is implied. Note that $\widehat E^A_j$ represents the ideal correlation event at the stage at which the record is formed; the bare macroscopic pointer event is $\widehat I_S\otimes|\alpha_j\rangle\langle\alpha_j|$. On the correlated subspace produced by an ideal premeasurement the two yield identical record statistics, although only the pointer component need remain stable under subsequent interactions with the environment. Two further properties of the record events should be noted. First, the elementary records do not add up to the certain event, $\sum_j\widehat E^A_j\ne\widehat I$; in accordance with Postulate 2, the sample space is completed by the complementary null outcome (``no record''). Second, while the bare pointer readings of two macroscopic devices commute, the record--correlation projectors associated with different bases of $\mathcal S$ do not: their incompatibility resides entirely in the $\mathcal S$ factor.

Conditioning on a record is what replaces the state preparation. Indeed, for an arbitrary subsequent event $\omega$, represented by a projector $\widehat P(\omega)$, the prediction formula, Eq.~(\ref{predret}), applied to the history $(a_j,\omega)$ gives
\begin{equation}
\label{Lueders}
\mathbb P(\omega|a_j)=\frac{\Tr\left[\widehat E^A_j\widehat P(\omega)\widehat E^A_j\right]}{\Tr\,\widehat E^A_j}=\Tr\left[\widehat P(\omega)\,\widehat\rho_j\right],
\qquad
\widehat\rho_j=\frac{\widehat E^A_j}{\Tr\widehat E^A_j},
\end{equation}
in accordance with Eq.~(\ref{rhoh}). In other words, conditioning on $a_j$ prepares the state $\widehat\rho_j$ ``out of unity'': the System is ``collapsed'' onto $|\psi_j\rangle$, while the rest of the world remains uniform, the least biased assignment compatible with the record, in line with the Maximum Entropy reading of Postulate 3. More generally, conditioning on an additional event updates the density operator according to L\"uders' rule \cite{lueders1951}, $\widehat\rho\rightarrow\widehat P\widehat\rho\widehat P/\Tr[\widehat P\widehat\rho\widehat P]$, of which Eq.~(\ref{Lueders}) is the special case of the uniform prior. This is precisely the content of von Neumann's projection postulate, obtained here as a theorem about conditional probabilities. The ideal, projective character of the update is inherited from the chain-operator structure of Postulate 3, which multiplies the history by the projector of each recorded event; more general measurements, described by quantum instruments, can be accommodated in the standard way, by dilating them to projective observations on a suitably extended system. Crucially, no non-unitary physical process is involved: the ``collapse'' is nothing but the Bayesian updating of the probability assignment upon conditioning on an additional event.

By Eq.~(\ref{Lueders}), for any subsequent event of the System itself the probabilities coincide with the abstract Born formula, Eq.~(\ref{BornBB}): the record faithfully stands in for the event it registers. In particular, a repeated ideal measurement of the same observable is guaranteed to reproduce the recorded outcome. If the subsequent event is itself a record of another device, its probability acquires an extra prior factor: under the uniform prior, it is unlikely that a second device is correlated with $\mathcal S$ at all. That factor is removed either by conditioning on the very existence of the second record or, dynamically, by preparing the device in a low-entropy ready state, which in the present framework is simply one more conditioning event of the history (Sec.~\ref{sec:spacetime}). The formulation thus does not eliminate initial conditions; it represents them as events, rather than as an externally supplied density operator.

Note that the Heisenberg cut has thereby lost its fundamental status: the boundary between ``the measured'' and ``the measuring'' can be drawn anywhere, or nowhere, without affecting the probabilities of the recorded events. Note also that the definite-outcome problem does not arise here in its standard form: since the postulates never refer to the wave function, there is no superposed fundamental state from which a unique actual outcome would have to be selected. The occurrence of the events themselves is primitive, governed by the Born--Boltzmann probabilities. What remains to be explained is why the pointer events of macroscopic apparatuses behave as stable, mutually exclusive classical alternatives. As discussed in Sec.~\ref{sec:maxent_decoherence}, this is precisely the question answered by environment-induced decoherence; and since Born's rule has now been obtained without invoking the WFC, the decoherence argument is no longer circular.

\subsection{Formal equivalence with standard sequential QM}
\label{sec:equivalence}
As an external consistency check, the logical relation can also be run in reverse: standard sequential Born--L\"uders quantum mechanics, evaluated on the finite-dimensional maximum-entropy prior, reproduces the Born--Boltzmann rule. Assume the textbook rules: (i) a measurement of an observable characterized by the PDI $\{\widehat P(\omega)\}$, performed on a system in a state $\widehat\rho$, yields the outcome $\omega$ with the probability $\Tr[\widehat P(\omega)\widehat\rho\,]$ (Born's rule); (ii) upon that outcome, the state is updated according to the projection postulate, $\widehat\rho\rightarrow\widehat P(\omega)\widehat\rho\,\widehat P(\omega)/\Tr[\widehat P(\omega)\widehat\rho\,]$; and (iii) in the absence of any prior information, the state of the system is given by the least biased, maximum-entropy density operator,
\begin{equation}
\label{maxentprior}
\widehat\rho_0=\frac{\widehat I}{D},\qquad D=\dim\mathcal H.
\end{equation}
Iterating (i) and (ii) for a sequence of consecutive ideal measurements with the outcomes $\omega_1,\dots,\omega_n$ then yields the standard sequential-measurement probability,
\begin{align}
p(\omega_1,\dots,\omega_n)&=\Tr\left[\widehat P(\omega_n)\cdots\widehat P(\omega_1)\,\widehat\rho_0\,\widehat P(\omega_1)\cdots\widehat P(\omega_n)\right]\nonumber\\
&=\Tr\left[\widehat C\,\widehat\rho_0\,\widehat C^\dagger\right]=\frac{W(\vec\omega^{(n)})}{D}.
\label{sequential}
\end{align}
All the joint probabilities of the resolved event sequences are thus proportional to the weights of Eq.~(\ref{weight}); in particular, Eq.~(\ref{probrule}) is recovered by applying Bayes' theorem to the binary refinement $\{\omega_k,\bar\omega_k\}$ at the $k$-th step, and Eq.~(\ref{blockcond}) to the sequences of steps. (On the standard side, the coherent coarse history $(L,R)$ and its resolved refinement correspond to the $k$-th measurement not being performed and being performed, respectively.) The two formulations are therefore formally equivalent, in the sense that they assign identical weights and conditional probabilities to the same specified histories and refinements of histories. What distinguishes them is the choice of the primitive notions. The standard formulation starts from the evolving quantum states, the WFC, and an externally supplied density operator; the present one starts from the events and a single probability rule, with the states, the collapse, and the preparation recovered as derived constructs. In particular, this equivalence provides a compact characterization of the Born--Boltzmann weight: it is, in finite dimensions, the standard sequential (L\"uders) probability evaluated on the maximum-entropy prior, Eq.~(\ref{maxentprior}), i.e. $p(h)=W(h)/D$.

\subsection{Time, space, and unitary evolution}
\label{sec:spacetime}
Note that no dynamics has been invoked so far: Born's rule, the projection postulate, and the density-operator formalism all follow from the timeless core, Postulates 1--3. To describe the evolution of physical systems, the core is supplemented with a statement about the symmetries of an isolated system:

\begin{enumerate}
\setcounter{enumi}{3}
\item {\bf Spacetime Translations:} {\it Translations of an isolated system in time and in space are represented by deterministic unitary transformations of its state space, forming continuous groups: a one-parameter group $\widehat U(t)$ and a three-parameter group $\widehat U(\mathbf x)$.}
\end{enumerate}
By Stone's theorem \cite{stone1932}, applied to each one-parameter subgroup, the corresponding generators are Hermitian operators,
\begin{equation}
\label{Stone}
\widehat U(t)=\exp\left(-\frac{i}{\hbar}\widehat H t\right),\qquad
\widehat U(\mathbf x)=\exp\left(-\frac{i}{\hbar}\widehat{\mathbf P}\cdot\mathbf x\right),
\end{equation}
identified with the Hamiltonian $\widehat H$ and the momentum $\widehat{\mathbf P}$, respectively. Note that Postulate 4 places time and space on an equal footing: the Hamiltonian plays for translations in time exactly the role that momentum plays for translations in space. This also clarifies the sense in which the core of the theory is ``timeless'': once the time labels are available, a physical temporal history is an ordered sequence of Heisenberg-picture events, with the ordinal order of Postulate 3 chosen consistently with the temporal (causal) order. The claim is not that physical time is derived from the event ordering, but that the probabilistic core of the theory requires nothing beyond the ordering itself. This symmetry-based treatment suggests a natural route toward a relativistic extension: for a relativistic system, $\widehat H/c$ and $\widehat{\mathbf P}$ combine into the four-momentum operator, and Postulate 4 is upgraded to the statement that an isolated system carries a unitary representation of the Poincar\'e group. A complete relativistic treatment, involving local algebras of events associated with spacetime regions and their consistency at spacelike separations, is left for future work; we only note that, since Postulates 1--3 never single out a time variable or a preferred foliation, nothing in the probabilistic core of the theory obstructs such an extension.

Since the primitive objects of the present formulation are the events rather than the states, the theory is closer in spirit to the Heisenberg picture than to the Schr\"odinger one, and the natural action of the translations is on the event algebra. An observation displaced in time by $t$ (or in space by $\mathbf x$) is represented by the conjugated projector,
\begin{equation}
\label{HeisP}
\widehat P(t)=\widehat U^\dagger(t)\,\widehat P\,\widehat U(t),
\end{equation}
and similarly for any operator built out of the events, $\widehat A(t)=\widehat U^\dagger(t)\widehat A\widehat U(t)$. Differentiation then yields the Heisenberg equation of motion,
\begin{equation}
\label{HeisEOM}
i\hbar\,\frac{d\widehat A(t)}{dt}=\left[\widehat A(t),\widehat H\right].
\end{equation}
In this, Heisenberg, picture the density operator, i.e. the preparation information, is static, while the projectors representing the events carry the dynamics. The familiar Schr\"odinger picture arises as the equivalent description in which the translation is instead absorbed into the preparation event: the operators do not evolve, while $\widehat \rho(t)=\widehat U(t) \widehat\rho (0)\widehat U^\dagger(t)$, and each pure component of the density operator obeys the Schr\"odinger equation~(\ref{Schro}) for the respective wave function $|\Psi_i\rangle$. Finally, the measurement records introduced in Sec.~\ref{sec:born} are produced by genuine unitary pre-measurement processes, Eq.~(\ref{premeasure}), generated by suitable interaction Hamiltonians, exactly as in von Neumann's original scheme. In particular, when the ready state of the apparatus, $|\alpha_0\rangle$ in Eq.~(\ref{premeasure}), is included as an additional conditioning event of the history, the pre-measurement interaction correlates the apparatus with the System deterministically, so that the record exists with certainty and the prior factor for the record's existence, noted in Sec.~\ref{sec:born}, is replaced by unity.

\section{Decoherence and the quantum-to-classical transition}
\label{sec:maxent_decoherence}
With the probabilistic rules now postulated without any reference to the WFC or to a classical observer, the formulation supplies an independent probabilistic foundation for the partial trace and the reduced density operators employed by the decoherence program: decoherence is no longer asked to justify the probability rule that is used to interpret its own reduced states, which was the circularity discussed in Sec.~\ref{sec:notenough}. In particular, the maximum-entropy characterization of the decohered state, Eq.~(\ref{maxent_var}), and its reading as a special case of the second law of thermodynamics acquire a firm foundation: the partial trace behind the reduced density operator, and the probabilistic meaning of its matrix elements, are now consequences of Postulate 3 rather than independent assumptions.

For the histories framework, decoherence addresses the remaining gap. The additivity caveat of Sec.~\ref{sec:BB} concerns the interference functional of Eq.~(\ref{interference}): the weight of a coherent coarse history differs from the sum over its refinement by the off-diagonal terms $D(h_i,h_j)$. Decoherence suppresses precisely these terms for the histories of macroscopic records: once $D(h_i,h_j)\simeq 0$ for $i\neq j$, the coherent coarse-graining becomes operationally indistinguishable from the classical, additive one, and the pointer events form, for all practical purposes, a classical sample space. (Note that decoherence does not make non-commuting projectors commute; it suppresses the interference between the relevant histories.) The measurement records of Sec.~\ref{sec:born} are thus precisely the structures that the dynamics itself renders stable and mutually exclusive.

The Maximum Entropy Principle thus enters the theory at three distinct levels: as the uniform prior of Postulate 3, as the preparation rule induced by the prehistory, Eq.~(\ref{rhoh}), and, as shown in Sec.~\ref{sec:EID}, as a dynamical attractor, with the emergent Process 1 a special case of the second law of thermodynamics. Within this model, then, the quantum-to-classical transition is the Maximum Entropy Principle at work.

The picture of Fig.~\ref{fig:cut} can now be completed. The Environment endows the Apparatus with emerging classicality, as in panel (c); and since the classical domain itself is no longer a primitive of the theory, the dashed boundary can at last be erased. What remains is a single quantum world, described by the postulates of Sec.~\ref{sec:timeless}, in which classicality, i.e. stable, mutually exclusive, additive records, is the emergent, maximum-entropy description of macroscopic subsystems.

\section{Discussion}
\label{sec:discussion}
Let us summarize the logical structure of the proposed formulation. The probabilistic core of QM consists of three postulates: the Hilbert state space, the representation of events by projection operators, and the Born--Boltzmann probability rule assigning weights to histories of events. The last of these combines, in a single axiom, Born's probability structure with the Maximum Entropy Principle: in the classical limit of compatible events it reduces to Boltzmann's counting of equally probable microstates, and its trace structure implements the uniform, least biased prior over the state space. Neither the WFC nor the wave function itself appears among the fundamental notions: the primitive elements are the events, whose occurrence is itself primitive; this avoids the standard formulation of the definite-outcome problem, the selection of a single branch of a universally evolving wave function. The quantum states are recovered as derived, informational objects, with the collapse reduced to L\"uders' Bayesian updating, and the microcanonical ensemble of statistical mechanics becomes a corollary rather than an extra assumption. The resulting theory is formally equivalent to the standard formulation, once the same histories and refinements are specified: each can be recovered from the other, the difference residing in the choice of the primitive notions. Dynamics is attached to this core by one additional postulate, which associates unitary transformations with translations in time and space, generated by the Hamiltonian and the momentum operators; time and space thus enter the theory on an equal footing, which suggests a natural route toward a relativistic extension. Finally, environment-induced decoherence, invoked without circularity, acts as a dynamical enforcer of the same Maximum Entropy Principle, in effect a special case of the second law of thermodynamics, driving macroscopic apparatuses to stable classical records and removing the need for the Heisenberg cut.

The construction owes most to the Consistent and Decoherent Histories programs \cite{hist_grif, hist_omnes, Hartle, decoh_hist, hist_Isham, Tkachenko2020}, from which it inherits its central object, the chain operator, and to the decoherence program \cite{zeh1970, Zurek_PT, Zurek_RMP, schlosshauer2007decoherence}, which supplies its dynamical content. It differs from the former in three respects: no consistency condition is imposed (the conditional probabilities of the Born--Boltzmann rule are normalized by construction, with the suppression of interference for macroscopic alternatives delegated to decoherence); the histories carry no time labels, the ordering of events being the primitive notion; and the probability rule is anchored in the uniform maximum-entropy prior: no density operator is supplied as an independent primitive, and the preparation information is specified through conditioning events. Within the histories family, the closest relative at the level of formulas is the history-operator formulation of Castellani \cite{castellani2019}, which employs essentially the same forward-conditioning rule; the present approach differs in the Maximum-Entropy interpretation of the trace, the record-projector treatment of measurements, and the replacement of the initial state by conditioning events. The derivations of Born's rule from decision theory or from envariance \cite{deutsch1999quantum, ManyWorlds, Zurek_PRA2005, Zurek2018, SebensCarroll} pursue a more ambitious goal, and the debate about their assumptions \cite{Barnum2000, envience2023noBorne} is a debate about how much can be obtained from how little. The present approach is deliberately more modest: it does not attempt to derive Born's rule at all; its probabilistic content is incorporated, together with the equal-a-priori-probability postulate of statistical mechanics, into the single Born--Boltzmann axiom. Gleason-type and reconstruction theorems \cite{Gleason1957, galley2017classification, Masanes2019} may be regarded as an independent motivation for this particular choice of the probability measure, although a uniqueness theorem tailored to the histories setting remains an open problem; in any case, probabilities here refer to objective, observer-independent records, and the WFC is never invoked. The Bayesian reading of the collapse resonates with epistemic interpretations of the quantum state, with an important difference: the events, and the weights assigned to them, are objective features of the world rather than degrees of belief of an agent.

In summary, we have argued that the wave function collapse can be removed from the postulates of quantum mechanics, with the measurement content absorbed into a single Born--Boltzmann probability rule of a transparent statistical-mechanical meaning. In the resulting formulation, the wave function, the classical observer, and the Heisenberg cut are not fundamental ingredients of the theory, but features of its derived, coarse-grained description.

\section*{Acknowledgments} This research has been performed at the Center for Functional Nanomaterials (CFN), which is a U.S. Department of Energy Office of Science User Facility, at Brookhaven National Laboratory under Contract No. DE-SC0012704.

\bibliographystyle{unsrtdoi}
\bibliography{q_lib_final}

\end{document}